# *Single-Interface Casimir Torque*

Tiago A. Morgado[1], Mário G. Silveirinha[1,2*]

[1]*Department of Electrical Engineering, Instituto de Telecomunicações, University of Coimbra, 3030-290 Coimbra, Portugal*

[2]*University of Lisbon, Instituto Superior Técnico, Avenida Rovisco Pais, 1, 1049-001 Lisboa, Portugal*

*E-mail:* tiago.morgado@co.it.pt, mario.silveirinha@co.it.pt

## Abstract

It is shown that the quantum fluctuations of the electromagnetic field generally induce a torque at the interface of an anisotropic material with another anisotropic or isotropic material. It is proven that this torque depends on an interface zero-point energy determined by the dispersion of the interface (localized and extended) modes. Our theory demonstrates that the single-interface torque is essential to understand the Casimir physics of material systems with anisotropic elements and determines the equilibrium positions of the system.

---

[*] To whom correspondence should be addressed: E-mail: mario.silveirinha@co.it.pt

# I. INTRODUCTION

Casimir-Lifshitz interactions [1-3] are the most paradigmatic example of quantum effects on the macro scale, and result from the confinement of the quantum-mechanical zero-point fluctuations of the electromagnetic field. Until recently, the study of quantum fluctuation-induced electromagnetic interactions was only of pure theoretical interest. Nevertheless, with the rapid development of micro- and nano-electromechanical systems (MEMS and NEMS) and its great impact in different areas [4-5], the research of Casimir-Lifshitz interactions has become of great practical importance as well. If, on one hand, Casimir interaction phenomena may lead to potentially undesired effects such as 'stiction' [6-7], on the other hand, they may open new and exciting possibilities in the field of micro and nanomechanics [4, 8-11].

The study of Casimir-Lifshitz phenomena was pioneered by Casimir for more than 60 years ago [1]. In his seminal work, Casimir showed that as a result of the electromagnetic field quantum fluctuations, two parallel perfectly conducting plates standing in a vacuum may experience an attractive force pushing the plates toward each other. Following Casimir's prediction, Lifshitz, Dzyaloshinskii, and Pitaevskii extended the theory to the more general case of realistic isotropic dielectric plates (including non-ideal metals) [2-3]. Some years later, this theory was further generalized to anisotropic dielectric plates [12-13]. Interestingly, it was shown that the anisotropy may lead to the emergence of qualitatively different phenomena. It was demonstrated that a pair of parallel anisotropic uniaxial plates (with in-plane optical anisotropy and misaligned optical axes) separated by an isotropic dielectric, may experience a mechanical torque, designated as Casimir torque, that spontaneously forces the rotation of the plates towards the minimum energy position (with the two optical axes aligned). The Casimir torque in this kind of systems was further investigated in [14-18]. In particular, numerical calculations of the torque were provided in [14-16, 18], and possible experiments to measure the Casimir torque were proposed in [14, 16-18].

With the emergence of metamaterials and their intriguing electromagnetic properties, the study of the Casimir-Lifshitz interactions has been also extended to systems involving complex structural nanoscopic unities [19-25]. In particular, the Casimir-Lifshitz phenomena have been investigated in systems formed by dense arrays of nanowires [26-27]. It was demonstrated in [26] that interactions mediated by nanowire-based materials result in ultralong-range forces, contrasting with the short-range Casimir forces characteristic of interactions mediated by isotropic dielectrics. The anomalous long-range Casimir force stems from the ultra-large density of photonic states in the nanowire materials, which boosts the quantum fluctuation-induced interactions [26-27].

Furthermore, in a recent work [28], we studied the Casimir interaction torque in nanowire materials, and demonstrated that it is distinctively different from the torques studied hitherto in other systems (e.g., birefringent parallel plates [14]). On one hand, it was proven that the Casimir interaction torque in nanowire structures has an unusual scaling law. Specifically, the torque generated due to the coupling between two interfaces decays as $1/d$ at large distances ($d$ is the distance between the two interfaces), which differs markedly from the characteristic $1/d^3$ decay in usual configurations wherein the two interfaces are separated by an isotropic background [28]. On the other hand, it was argued that the torque has an additional and dominant contribution, designated by Casimir single-interface torque, which is an interfacial effect and does not vanish even when the two interfaces are infinitely far apart. The study of [28] was however mainly qualitative, and no detailed quantitative analysis of the single-interface torque was provided. The objective of this work is to study in depth this single-interface torque and unveil the physical mechanisms associated with this nontrivial Casimir-type interaction.

Even though the analysis in [28] was focused on nanowire materials, the single-interface torque emerges at any interface involving at least an anisotropic material with optical axes out of the interface plane. In these conditions, the zero-point energy of the system depends on the relative orientation of the material optical axes. Thus, rather than considering the particular case of metallic nanowire systems, here we theoretically investigate the Casimir single-interface torque in general conditions, treating the relevant anisotropic materials as continuous media.

## II. MICROSCOPIC THEORY

### A. Zero-point energy

We are interested in the Casimir-type interactions between different anisotropic materials at zero-temperature. Even though at a later stage the relevant media will be modeled as continuous anisotropic uniaxial dielectrics, in a first step it is convenient to visualize each material as a periodic arrangement of inclusions embedded in a vacuum (Fig. 1) and develop the theory relying on such a microscopic model. The inclusions may be pictured as either spherical or ellipsoidal depending if the material response is isotropic or anisotropic. For each material region the optical axis is assumed to be in the *yoz* plane and we define $\hat{\mathbf{u}}_\alpha = \sin\alpha \hat{\mathbf{u}}_y + \cos\alpha \hat{\mathbf{u}}_z$ as the unit vector oriented along the optical axis. The angle $\alpha$ determines the orientation of the inclusions in the pertinent material region.

The zero-point energy $\varepsilon_C$ of the system can be calculated with the help of the argument principle [29-32]. In this section, we consider a generic double-interface configuration (Fig. 1(b)) and revisit the usual derivation of the zero-temperature Casimir energy [29-32]. We start by noting that if $D(\omega, \mathbf{k}_\parallel, \boldsymbol{\alpha}, d) = 0$ represents the characteristic equation of the photonic modes with transverse wave vector $\mathbf{k}_\parallel = (k_x, k_y)$, the argument principle implies that:

$$\sum_m \frac{\hbar}{2}\omega^Z_{\mathbf{k},m} - \sum_m \frac{\hbar}{2}\omega^P_{\mathbf{k},m} = \frac{1}{2\pi i} \oint_C \frac{\hbar\omega}{2} \frac{\partial_\omega D}{D} d\omega, \qquad (1)$$

where $\hbar = h/(2\pi)$ is the reduced Planck constant, $\boldsymbol{\alpha} = (\alpha_1, \alpha_2, \alpha_3)$, $\omega_{\mathbf{k},m}^Z$ represents a generic zero of $D$ inside the closed contour $C$ and $\omega_{\mathbf{k},m}^P$ represents a generic pole of $D$. When the middle region is a vacuum – as assumed in this section – the angle $\alpha_2$ has no meaning and can be ignored. Yet, we will keep it in the formulas because at a later stage we will consider the general case where the middle region is an anisotropic material.

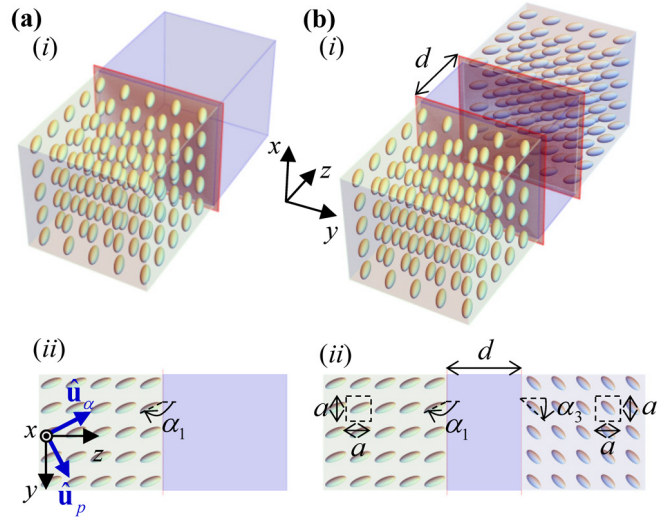

**Fig. 1.** Sketch of the system under study. (a) Single-interface configuration: anisotropic dielectric-vacuum interface. (b) Double-interface configuration: anisotropic dielectric I – vacuum – anisotropic dielectric II. The thickness of the vacuum region is $d$.

Generalizing the approach of Ref. [32] to three-dimensional geometries, it follows that for a periodic system the characteristic function $D$ may be chosen of the form

$$D(\omega, \mathbf{k}_\parallel, \boldsymbol{\alpha}, d) \equiv \det\left[\underline{\mathbf{1}} - \underline{\mathbf{R}}_\mathrm{L}(\omega, \mathbf{k}_\parallel, \alpha_1, \alpha_2) \cdot \underline{\mathbf{M}}_\mathrm{B}(\omega, \mathbf{k}_\parallel, \alpha_2, d) \cdot \underline{\mathbf{R}}_\mathrm{R}(\omega, \mathbf{k}_\parallel, \alpha_2, \alpha_3) \cdot \underline{\mathbf{M}}_\mathrm{F}(\omega, \mathbf{k}_\parallel, \alpha_2, d)\right] \quad (2)$$

where $\underline{\mathbf{1}}$ is a unit matrix, $\underline{\mathbf{R}}_\mathrm{L,R}$ are the reflection matrices for the left and right interfaces, and $\underline{\mathbf{M}}_\mathrm{F,B}$ are the propagation matrices for the forward waves (travelling along the $+z$ direction in region 2) and the backward waves (travelling along the $-z$ direction in region 2) [see Fig. 1(b)]. The associated basis of functions is formed by the vacuum plane wave modes (both propagating and evanescent) with transverse wave vector of the form $\mathbf{k}_\parallel + \mathbf{G}$ ($\mathbf{G}$ is a generic transverse reciprocal lattice vector) which can be used to expand a generic wave with the Bloch property in the transverse (*x* and *y*) coordinates [32]. The matrices $\underline{\mathbf{R}}_{L,R}$ and $\underline{\mathbf{M}}_{F,B}$ have infinite dimension, and the transverse wave vector must be restricted to the 1st Brillouin zone (B.Z.) [23, 32].

Summing both members of Eq. (1) over all possible wave vectors, it is possible to write

$$\frac{1}{A}\sum_{\mathbf{k},m}\frac{\hbar}{2}\omega_{\mathbf{k},m}^{Z} - \frac{1}{A}\sum_{\mathbf{k},m}\frac{\hbar}{2}\omega_{\mathbf{k},m}^{P} = \frac{1}{2\pi i}\frac{1}{(2\pi)^2}\iint_{B.Z.} dk_x dk_y \oint_C \frac{\hbar\omega}{2}\frac{\partial_\omega D}{D}d\omega, \quad (3)$$

where $A = L_x \times L_y$ is the cross-sectional area of the cavity parallel to the *xoy* plane. As usual, *C* is taken as a contour oriented counter-clockwise that consists of the imaginary frequency axis, and of a semi-circle with infinite radius in the semi-plane $\text{Re}\{\omega\} > 0$. Assuming that the material response ceases when $\omega \to \infty$ it follows that $D(\omega, \mathbf{k}_\parallel, \boldsymbol{\alpha}, d)$ becomes independent of both *d* and **α** when $\omega \to \infty$, and thus the integral over the semi-circle is a constant independent of the system configuration and may be dropped. Moreover, noting that the first term in the left-hand side of Eq. (3) is the zero-point energy per unit of area, we can write:

$$\frac{1}{A}\varepsilon_{C,\text{tot}}(d,\boldsymbol{\alpha}) - \frac{1}{A}\sum_{\mathbf{k},m}\frac{\hbar}{2}\omega_{\mathbf{k},m}^{P} = \frac{1}{2\pi i}\frac{1}{(2\pi)^2}\iint_{B.Z.} dk_x dk_y \int_{i\infty}^{-i\infty}\frac{\hbar\omega}{2}\frac{\partial_\omega D}{D}d\omega. \quad (4)$$

After integration by parts, the right-hand side of this formula reduces to the familiar Casimir interaction energy defined as:

$$\frac{\delta\varepsilon_{C,\text{int}}}{A} = \frac{\hbar}{4\pi^3}\int_0^{\pi/a} dk_x \int_{-\pi/a}^{\pi/a} dk_y \int_0^{+\infty} d\xi \log D(i\xi, k_x, k_y, \boldsymbol{\alpha}, d). \quad (5)$$

where $\xi$ is the imaginary frequency ($\omega = i\xi$) and *a* is the lattice period. We used the fact that $B.Z. = [-\pi/a, \pi/a] \times [-\pi/a, \pi/a]$ and that *D* is an even function of $k_x$. Thus, we have proven that:

$$\frac{1}{A}\varepsilon_{C,\text{tot}}(d,\boldsymbol{\alpha}) = \frac{1}{A}\delta\varepsilon_{C,\text{int}}(d,\boldsymbol{\alpha}) + \frac{1}{A}\sum_{\mathbf{k},m}\frac{\hbar}{2}\omega_{\mathbf{k},m}^{P}. \quad (6)$$

One crucial point is that the poles $\omega^P_{\mathbf{k},m}$ of $D$ must be independent of $d$. This is why the second term in the right-hand side of Eq. (6) can be disregarded in the calculation of the Casimir force. In our formulation the poles $\omega^P_{\mathbf{k},m}$ correspond to the poles of the reflection coefficients $\underline{\mathbf{R}}_L$ and $\underline{\mathbf{R}}_R$ associated with the two individual material interfaces, which are evidently independent of $d$ but which depend on $\boldsymbol{\alpha}$. This property shows that the second term in the right-hand side of Eq. (6) can be decomposed as:

$$\frac{1}{A}\sum_{\mathbf{k},m}\frac{\hbar}{2}\omega^P_{\mathbf{k},m} = \frac{1}{A}\varepsilon_{C,12}(\alpha_1,\alpha_2) + \frac{1}{A}\varepsilon_{C,23}(\alpha_2,\alpha_3), \tag{7}$$

where $\varepsilon_{C,12}(\alpha_1,\alpha_2)$ ($\varepsilon_{C,23}(\alpha_2,\alpha_3)$) represents $\sum_{\mathbf{k},m}\frac{\hbar}{2}\omega^P_{\mathbf{k},m}$ with the summation range restricted to the poles of $\underline{\mathbf{R}}_L$ ($\underline{\mathbf{R}}_R$). As is well-known, the poles of the reflection coefficients correspond to the guided modes supported by the individual interfaces. Thus, the left-hand side of Eq. (7) has a clear physical meaning: it is the zero-point energy associated with the edge modes supported by the two uncoupled interfaces. In other words, $\varepsilon_{C,12}$ and $\varepsilon_{C,23}$ in Eq. (7) correspond to the zero-point energies of the guided modes supported by each interface. One important aspect is that the spatial domain is required to be electromagnetically closed. Hence, the cavity should be terminated with some type of opaque boundary, for example with periodic boundary conditions or a perfectly electric conducting wall placed at $z = \pm\infty$. Thus, strictly speaking the poles of $\underline{\mathbf{R}}_L$ and $\underline{\mathbf{R}}_R$ do not need to be associated with waves localized at the interfaces, and may be associated with spatially extended modes.

In summary, we formally demonstrated that when the materials response ceases for $\omega \to \infty$ the zero-point energy of the double-interface configuration (Fig. 1(b)) can be written as (apart from an irrelevant constant independent of the system configuration):

$$\varepsilon_{C,\text{tot}}(d,\boldsymbol{\alpha}) = \delta\varepsilon_{C,\text{int}}(d,\boldsymbol{\alpha}) + \varepsilon_{C,12}(\alpha_1,\alpha_2) + \varepsilon_{C,23}(\alpha_2,\alpha_3). \tag{8}$$

The first term $\delta\varepsilon_{C,int}$ corresponds to the usual Casimir interaction energy that appears due to the coupling between the two interfaces, whereas the other two terms are associated with the Casimir single-interface energies determined by the orientation of the optical axes. These single-interface components are due to the anisotropy of the materials because the energy of the system depends on the angles $\boldsymbol{\alpha} = (\alpha_1, \alpha_2, \alpha_3)$ that dictate the orientation of the inclusions. Even though the single-interface terms $\varepsilon_{C,12}$ and $\varepsilon_{C,23}$ are distance independent, and therefore do not contribute to the usual Casimir force, they can contribute to the Casimir torque. This will be discussed in detail in the next subsection.

## B. Casimir torque

Next, we derive the Casimir torque in the considered material structure, and highlight the differences compared to the torques induced in conventional systems with in-plane anisotropy.

The total Casimir torque acting on the *i*-th body (*i*=1,2,3) in the double-interface configuration is $M_{C,tot}^{(i)} = -\partial \varepsilon_{C,tot}/\partial \alpha_i$, and hence from Eq. (8) it is given by:

$$M_{C,tot}^{(i)} = M_{C,int}^{(i)} + M_{C,12}^{(i)} + M_{C,23}^{(i)}$$
$$= -\frac{\partial \delta\varepsilon_{C,int}}{\partial \alpha_i} - \frac{\partial \varepsilon_{C,12}}{\partial \alpha_i} - \frac{\partial \varepsilon_{C,23}}{\partial \alpha_i}. \quad (9)$$

In systems where the interaction is mediated by an isotropic material and when the optical axes of the materials 1 and 3 are parallel to the interface, $\varepsilon_{C,12}$ and $\varepsilon_{C,23}$ are evidently independent of $\alpha_i$, and hence it is possible to assume that the Casimir zero-point energy of the system $\varepsilon_{C,tot}$ can be replaced by the interaction energy $\delta\varepsilon_{C,int}$. Thus, in such a scenario the Casimir torque is simply given by $M_{C,tot}^{(i)} = M_{C,int}^{(i)} = -\partial \delta\varepsilon_{C,int}/\partial \alpha_i$ [13], where $M_{C,int}^{(i)}$ is designated here by interaction torque. However, in a system where the optical axes of the relevant media are out of plane with respect to the interface this cannot be done. Indeed, in these conditions there are two additional contributions to the Casimir torque, namely $M_{C,12}^{(i)}$ and $M_{C,23}^{(i)}$. These two terms are designated here by single-interface torques and are independent of $d$. Clearly, when $d \to \infty$ the interaction torque vanishes $M_{C,\text{int}}^{(i)} = 0$ and $\lim_{d \to \infty} M_{C,\text{tot}}^{(i)} = M_{C,12}^{(i)} + M_{C,23}^{(i)}$. For example, for the 1$^{\text{st}}$ body one has $\lim_{d \to \infty} M_{C,\text{tot}}^{(1)} = M_{C,12}^{(1)}$ and for the 3$^{\text{rd}}$ body one has $\lim_{d \to \infty} M_{C,\text{tot}}^{(3)} = M_{C,23}^{(3)}$. Hence, $M_{C,12}^{(1)}$ and $M_{C,23}^{(3)}$ have a clear physical meaning: they are the individual torques induced at the interfaces 1-2 and 2-3, respectively, by the quantum fluctuations of the electromagnetic field.

A single-interface torque also occurs in the single-interface configuration (Fig. 1(a)). For such a configuration it is physically evident that there must be a preferred orientation for the optical axis of the medium, and hence some associated zero-point energy. To determine the single-interface energy $\varepsilon_{C,\text{s.i.}}$ and torque $M_{C,\text{s.i.}}$, we adapt the ideas of our previous work [28], and consider the scenario where the gap $d$ between the two interfaces in the double-interface configuration (Fig. 1(b)) is vanishingly small (i.e., $d = 0^+$). For clarity, let us consider a twin-interface scenario wherein the inclusions in region 1 and 3 are identical and $\alpha_1 = \alpha_3$. The limit $d = 0^+$ is understood here as the situation for which the regions 1 and 3 are merged to form a periodic (crystalline) structure, i.e. a bulk material. In this limit, the total Casimir energy may still depend on the orientation of the particles because even for a bulk crystal not all the directions of space are equivalent due to the granularity of the structure. Let us denote $M_{\text{bulk}}$ as the torque acting on the bulk crystal which depends on $\alpha_1 = \alpha_3$. Note that $M_{\text{bulk}}$ is expected to be proportional to the volume of the bulk crystal. Calculating the $d \to 0$ limit of both members of Eq. (8) and the derivative with respect to $\alpha_1 = \alpha_3$ we see that

$$M_{\text{bulk}}^{(1)} = -\frac{\partial}{\partial \alpha_1} \left[ \delta\varepsilon_{C,\text{int}}\left(d = 0^+, \alpha_1, \alpha_2, \alpha_1\right) + \varepsilon_{C,12}\left(\alpha_1, \alpha_2\right) + \varepsilon_{C,23}\left(\alpha_2, \alpha_1\right) \right]. \tag{10}$$

From here, we see that $M_{C,12}^{(1)} + M_{C,23}^{(3)} - M_{bulk}^{(1)} = \frac{\partial}{\partial \alpha_1}\left[\delta\varepsilon_{C,int}\left(d=0^+,\alpha_1,\alpha_2,\alpha_1\right)\right]$. But for a twin-material interface with $\alpha_1 = \alpha_3$ it is evident that $\varepsilon_{C,12}(\alpha_1,\alpha_2) = \varepsilon_{C,23}(\alpha_2,\alpha_1) \equiv \varepsilon_{C,s.i.}$ and hence $M_{C,12}^{(1)} = M_{C,23}^{(3)} \equiv M_{C,s.i.}^{(1)}$. Therefore, it follows that the single-interface Casimir torque is such that:

$$M_{C,s.i.}^{(1)} - \frac{1}{2}M_{bulk}^{(1)} = -\frac{1}{2}M_{C,int}\bigg|_{d=0^+} = -\frac{\partial \varepsilon_{C,s.i.}}{\partial \alpha_1}, \tag{11a}$$

$$\varepsilon_{C,s.i.} = -\frac{1}{2}\delta\varepsilon_{C,int,121}\bigg|_{d=0^+}, \tag{11b}$$

where $\delta\varepsilon_{C,int,121}$ is a short-hand notation for $\delta\varepsilon_{C,int}(d=0^+,\alpha_1,\alpha_2,\alpha_1)$. Evidently, $\varepsilon_{C,s.i.}$ is defined apart from the sum of an irrelevant constant. The above formulas give the single-interface energy and torque in terms of the interaction energy $\delta\varepsilon_{C,int}$ of a twin 1-2-1 configuration which can be calculated using Eq. (5). This derivation shows that the single-interface torque in general has a volumetric component $M_{bulk}^{(1)}/2$ and a surface correction (the right hand side of Eq. (11a). The factor $1/2$ is because $M_{C,s.i.}^{(1)}$ represents the torque acting on half of the crystalline structure. Thus, $M_{C,s.i.}^{(1)} - \frac{1}{2}M_{bulk}^{(1)}$ corresponds to the additional stress due to the asymmetry created by the interface, and consistent with this it is proportional to the area of the interface.

Even though the described theory is completely rigorous, the granularity of the crystal does not allow for a simple analytical treatment. To circumvent this issue, in the next section we consider the continuous medium approximation.

### III. MACROSCOPIC THEORY

*A. Continuum approximation*

It is possible to considerably simplify the problem using an effective medium approximation wherein each material region is seen as a uniaxial anisotropic dielectric with permittivity:

$$\overline{\overline{\varepsilon}} = \varepsilon_t \hat{\mathbf{u}}_x \hat{\mathbf{u}}_x + \varepsilon_t \hat{\mathbf{u}}_p \hat{\mathbf{u}}_p + \varepsilon_{\alpha\alpha} \hat{\mathbf{u}}_\alpha \hat{\mathbf{u}}_\alpha \tag{12}$$

where $\hat{\mathbf{u}}_p = \cos\alpha_i \hat{\mathbf{u}}_y - \sin\alpha_i \hat{\mathbf{u}}_z$ is a unit vector in the *yoz* plane perpendicular to the optical axis ($\hat{\mathbf{u}}_\alpha$). In the isotropic case (spherical inclusions) one has $\varepsilon_t = \varepsilon_{\alpha\alpha}$, whereas in the anisotropic case (elongated elliptical inclusions) $\varepsilon_t \neq \varepsilon_{\alpha\alpha}$.

In the continuum limit, for each fixed $\mathbf{k}_\parallel$ the electromagnetic fields in the vacuum region can be expanded simply in terms of the usual plane wave modes, similar to Ref. [28]. Hence, in this case the matrices $\underline{\mathbf{R}}_{L,R}$ and $\underline{\mathbf{M}}_{F,B}$ in Eq. (2) become 2×2 matrices and can be determined using standard analytical methods [28] (see also Appendix A). Indeed, within the effective medium framework the wave propagation is described by an ordinary wave (transverse electric (TE) mode) and an extraordinary wave (transverse magnetic (TM) mode) [28].

At this point, it is important to discuss the validity of the continuous medium approximation. Typically, effective medium methods are valid for interactions such that $k_\parallel a < 1$ and $\omega a/c < 1$. In the microscopic formulation $\delta\varepsilon_{C,\text{int}}$ must be calculated in the limit $d = 0^+$ for which the structure becomes periodic (a crystal). In this limit the distance between adjacent layers of inclusions is nonzero, but is as small as $\tilde{d} \approx a$, i.e., on the order of the lattice constant. Thus, it is possible to estimate that the modes relevant for the Casimir interaction satisfy $k_\parallel a < 1$ and $\omega a/c < 1$, which is precisely the limit of validity of the effective medium approximation. Due to this reason, it follows that the effective medium framework is only approximately satisfactory, and in particular it may not yield quantitatively precise results. Yet, the effective medium theory has the advantage that it enables a simple analysis of the problem, and we expect it to provide at least a qualitatively correct description of the physics of the single-interface Casimir torque.

Another important aspect is that in the continuum limit the torque in a bulk material must vanish ($M_{\text{bulk}} = 0$) because any orientation of the optical axis is energetically equivalent when there is no underlying granularity. Hence, in the continuum limit Eq. (11) becomes:

$$M_{\text{C,s.i.}}^{(1)} = -\frac{\partial}{\partial \alpha_1} \varepsilon_{\text{C,s.i.}}, \tag{13a}$$

$$\frac{\varepsilon_{\text{C,s.i.}}}{A} = -\frac{\hbar}{8\pi^3} \int_0^{\pi/a} dk_x \int_{-\pi/a}^{\pi/a} dk_y \int_0^{+\infty} d\xi \log D \big|_{d=0^+} \left( i\xi, \mathbf{k}_\parallel, \alpha_1, \alpha_2 \right), \tag{13b}$$

$$D\big|_{d=0^+}\left(i\xi, \mathbf{k}_\parallel, \alpha_1, \alpha_2\right) = \det\left[\underline{\mathbf{1}} - \underline{\mathbf{R}}_{\text{L}}(i\xi, \mathbf{k}_\parallel, \alpha_1, \alpha_2) \cdot \underline{\mathbf{R}}_{\text{R}}(i\xi, \mathbf{k}_\parallel, \alpha_2, \alpha_1)\right], \tag{13c}$$

so that the single-interface torque is only due to surface effects. We used the fact that in the limit $d = 0^+$ the propagation matrices $\underline{\mathbf{M}}_{\text{F,B}}$ become identical to the unit matrix. It is implicit that the double-interface structure corresponds to a twin-material configuration (1-2-1).

Note that in the continuum limit we let $d$ to be precisely zero in the calculation of the single-interface torque, but the transverse momentum is still restricted to the 1$^{\text{st}}$ Brillouin zone as in the periodic case. The justification for this is (*i*) the effective medium theory breaks down for $k_\parallel a > 1$, (*ii*) the wave vector cut-off $k_{\max} \sim \pi/a$ effectively mimics the fact that in the microscopic model the distance between the inclusions does not reach zero, but has a minimum on the order of $\tilde{d} = a$. Thus, only modes with $k_\parallel a < 1$ can effectively contribute to the single-interface Casimir torque.

It should be mentioned that without a wave vector cut-off (i.e., with $k_{\max} = \infty$) the integral in Eq. 13(b) would diverge because infinitely many photonic channels would contribute to the interaction. This result is unphysical because in the microscopic formalism the distance between adjacent planes of inclusions always exceeds $\tilde{d} \approx a$, and hence in the microscopic theory $\delta\varepsilon_{C,int}$ remains finite in the limit $d = 0^+$. The wave vector cut-off in the continuum approximation is essential so that the macroscopic theory can have the same features as the microscopic theory and predict a finite single-interface torque. It can be checked that the integral (13(b)) converges for $d = 0^+$ provided the effective dielectric response of the materials ceases for sufficiently high frequencies, i.e. that the dielectric permittivity of all relevant materials [Eq. (12)] approaches the vacuum permittivity when $\omega \to \infty$. This condition is always satisfied for realistic materials because the electric dipoles cannot follow very rapid oscillations of the electric field. In this situation the reflection matrices $\underline{\mathbf{R}}_i(i\xi)$ vanish when $\xi \to \infty$, and it can be checked that this implies that $\delta\varepsilon_{C,int}$ is finite.

In summary, the single-interface torque is originated by interactions between bodies that are nearly in contact ($\tilde{d} \approx a$) and hence an effective medium description of the problem depends critically on the high-frequency (both spatial and temporal) response of the materials. The precise knowledge of the effective dielectric function for $\omega \to \infty$ and the precise wave vector cut-off $k_{max}$ are critical to make quantitative predictions.

## *B. Generalization*

So far it was assumed that the middle layer (region 2 in Fig. 1) is a vacuum, so that $M_{C,s.i.}^{(1)}$ corresponds to the single-interface torque when the material 1 is adjacent to a vacuum. However, within the effective medium description there is no difficulty in generalizing the theory to the case wherein the middle layer is an arbitrary anisotropic dielectric (Fig. 2)

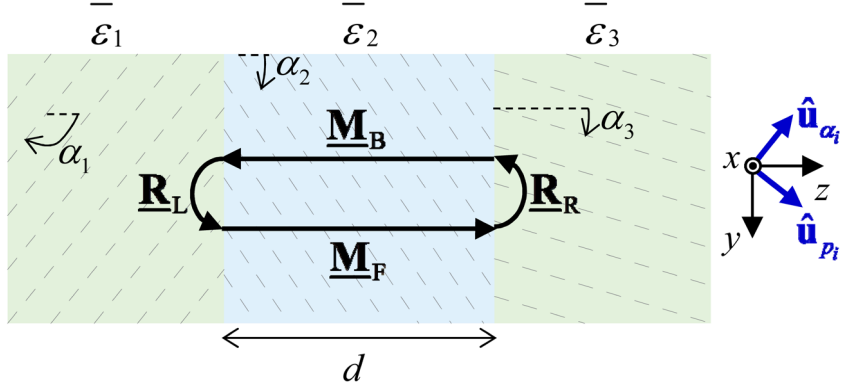

**Fig. 2.** Sketch of the double-interface configuration when the middle region is an arbitrary anisotropic dielectric and the relevant materials are regarded as a continuum. The gray dashed lines represent the optical axes of the materials. The zero-point energy of the system is calculated from the reflection $\underline{\mathbf{R}}_{L,R}$ and transfer $\underline{\mathbf{M}}_{F,B}$ matrices.

A straightforward analysis analogous to that reported in Sect. II, but using as a starting point the macroscopic framework with the physical cut-off $k_{max} \sim \pi/a$, shows that Eq. (13) remains valid when the middle region is an arbitrary dielectric. As before, $M_{C,s.i.}^{(1)} = -\frac{\partial}{\partial \alpha_1} \varepsilon_{C,s.i.}$ is understood as the single-interface torque acting on medium 1 for an interface between medium 1 and medium 2. However, when the second material is not isotropic the torque on medium 2 is typically nonzero, and can be calculated using $M_{C,s.i.}^{(2)} = -\frac{\partial}{\partial \alpha_2} \varepsilon_{C,s.i.}$.

It is important to prove that the theory is self-consistent. Indeed, $\varepsilon_{C,s.i.}$ in Eq. (13) is calculated by considering a twin configuration of the type 1-2-1 with thickness of the middle layer $d = 0^+$. However, in the macroscopic formulation there is no reason to regard the medium 1 as special as compared to medium 2. Indeed, one could alternatively calculate $\varepsilon_{C,s.i.}$ based on a 2-1-2 twin configuration where the middle layer has $d = 0^+$. Does this alternative calculation method yield the same Casimir energy ($\varepsilon_{C,s.i.,121} = \varepsilon_{C,s.i.,212}$)? The answer to the question is positive. Indeed, we prove in Appendices A and B [see Eq. (B3)] that the characteristic equations in the two scenarios are identical $D_{121}(\omega, k_x, k_y) = D_{212}(\omega, k_x, k_y)$ and consequently the single-interface Casimir energy is independent of the calculation method.

One can still imagine a different way to determine the torque for an interface of two materials. Let us now label the relevant materials as "1" and "3" and suppose that we want to calculate the torque on the material 1 for the single 1-3 interface ($M_{C,s.i.,13}^{(1)}$). As already discussed, one option is to use Eq. (13) for a twin-material configuration 1-3-1 (or alternatively 3-1-3): $M_{C,s.i.,13}^{(1)} = -\frac{\partial}{\partial \alpha_1} \varepsilon_{C,s.i.,131}$. Alternatively, one can consider instead a generic configuration 1-2-3 in the limit where the middle layer (region 2, which can be taken as an arbitrary material) has thickness $d = 0^+$. Using Eq. (9) and noting that in the limit $d = 0^+$ the torque $M_{C,tot}^{(1)}$ should be coincident (independent of the material in region 2) with $M_{C,s.i.,13}^{(1)}$ it is found that:

$$M_{C,s.i.,13}^{(1)} = -\frac{\partial}{\partial \alpha_1} \delta\varepsilon_{C,int,123}\bigg|_{d=0^+} - \frac{\partial}{\partial \alpha_1} \varepsilon_{C,12} - \frac{\partial}{\partial \alpha_1} \varepsilon_{C,23}$$
$$= -\frac{\partial}{\partial \alpha_1} \delta\varepsilon_{C,int,123}\bigg|_{d=0^+} - \frac{\partial}{\partial \alpha_1} \varepsilon_{C,s.i.,121} \quad (14)$$

The indices "123" and "121" identify the configuration used to evaluate the interaction energy and the single-interface energy, respectively. Does the above formula give the same result as $M_{C,s.i.,13}^{(1)} = -\frac{\partial}{\partial \alpha_1} \varepsilon_{C,s.i.,131}$? We will not attempt to give a direct proof of this property but in the next section it is shown with numerical simulations that the answer is affirmative. This result demonstrates that the theory is fully self-consistent, and that the calculated torque is, indeed, independent of the considered limit process.

## IV. NUMERICAL EXAMPLES

In order to characterize the single-interface energy and torque in the considered systems (Fig. 1), next we carry out extensive numerical simulations based on Eqs. (9) and (13). It is assumed that the anisotropic materials have $\varepsilon_t = 1$ and $\varepsilon_{\alpha\alpha} = \varepsilon_{\text{Lorentz}}$ such that $\varepsilon_{\text{Lorentz}}$ follows the Lorentz dispersion model

$$\varepsilon_{\text{Lorentz}} = 1 - \frac{\omega_e^2}{\omega^2 - \omega_0^2 + i\omega\Gamma}, \qquad (15)$$

where $\omega_0$ is the resonant frequency, $\omega_e$ determines the strength of the electric resonance, and $\Gamma$ is the damping factor related to material loss. For simplicity, the resonance frequency is taken equal to $\omega_0/(2\pi) = 95.49$ THz for all the materials. The parameter $\omega_e$ is material dependent. For convenience, we introduce the anisotropy ratio $\chi = \varepsilon_{\alpha\alpha}/\varepsilon_t$, which by definition is evaluated in the static limit ($\omega = 0$). In the simulations it was assumed that $\omega_0 a/c = 0.1$ (where $a$ is the lattice period and $c$ is the speed of light in vacuum) and $\Gamma = 0.05\omega_0$. In case of isotropic materials one has $\varepsilon_t = \varepsilon_{\alpha\alpha} = \varepsilon_{\text{Lorentz}}$.

### A. Single-interface configurations

To begin with, we study the single-interface Casimir interactions at the junction of an anisotropic and an isotropic material (Fig. 1(a)).

In the first example (Fig. 3), we consider a vacuum-anisotropic dielectric interface. The curves (*i*) of Figs. 3(a) and 3(b) show the calculated single-interface energy and the torque acting on the anisotropic material. As seen, the energy has a minimum when the optical axes of the anisotropic particles are parallel to the interface plane ($\alpha_2 = \pm 90°$). Such a configuration ensures that the dipoles in the last atomic layer (in the *yoz* plane) are aligned, which is a physically reasonable result. Thus, the quantum fluctuations lead to an internal surface stress that tends to orient the "elliptical-type" inclusions parallel to the interface. The configuration $\alpha_2 = \pm 90°$ corresponds to the stable equilibrium position. As to the single-interface torque, one can see from Fig. 3(b) (curve (*i*)) that it varies approximately as $\sin(2\alpha_2)$, somewhat analogous to the typical angle-dependence of the interaction torque but here the optical axis is not parallel to the interface [14].

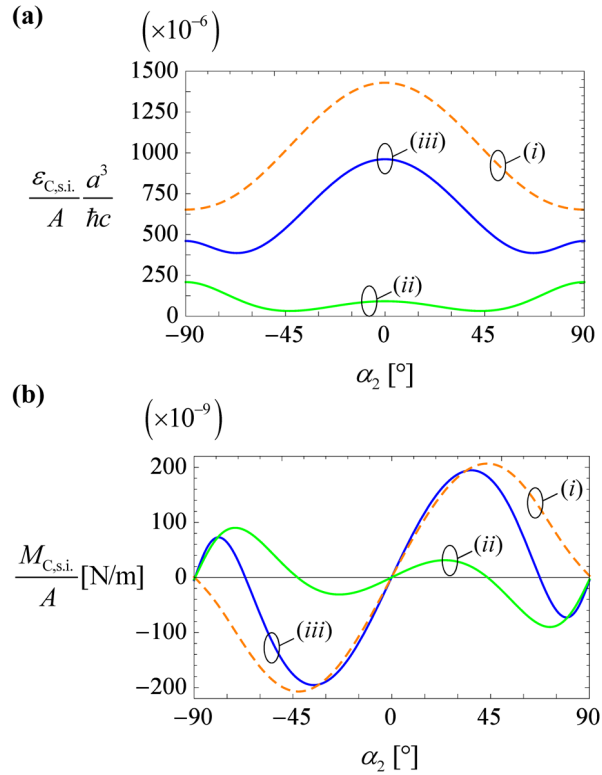

**Fig. 3.** (Color online) Normalized single-interface energy (a) and torque (b) at the junction between an isotropic dielectric (region 1) and an anisotropic dielectric (region 2), as a function of the angle $\alpha_2$. (*i*) Region 1: vacuum; Region 2: anisotropic material with anisotropy ratio $\chi = 5$ ($\omega_e / \omega_0 = 2$); (*ii*) Region 1: isotropic material with relative static permittivity $\varepsilon_{\omega=0} = 2$ ($\omega_e = \omega_0$); Region 2: anisotropic material with anisotropy ratio $\chi = 5$ ($\omega_e / \omega_0 = 2$)); (*iii*) Similar to (*ii*) but the anisotropic material has the anisotropy ratio $\chi = 10$ ($\omega_e / \omega_0 = 3$).

Interestingly, if the vacuum half-space is replaced by a dielectric material, e.g., with static permittivity $\varepsilon_{\omega=0} = 2$ (curves (*ii*)-(*iii*) of Fig. 3), the preferential orientation of the anisotropic particles is no longer parallel to the interface, and accordingly the single-interface torque does not exhibit a $\sin(2\alpha_2)$ variation. Specifically, one can see that when the anisotropy ratio is $\chi = 5$ (curves (*ii*)) the preferential orientation is $\alpha_2 = \pm 43°$, whereas when $\chi = 10$ (curves (*iii*)) it is $\alpha_2 = \pm 67°$. This effect can be understood noting that the electric dipoles in the isotropic region (which on average are expected to be randomly oriented in the bulk region) tend to attract the dipoles in the anisotropic material, leading in this way to a shift of the equilibrium position towards the normal direction.

Next, we study a configuration wherein the two juxtaposed semi-infinite materials are anisotropic. The two anisotropic materials have the same anisotropy ratio $\chi = 10$ but optical axes with different orientations. To begin with, we consider a scenario wherein the particles of the medium 1 have a fixed orientation $\alpha_1$, whereas the inclusions of medium 2 are free to conjointly rotate in the *yoz* plane.

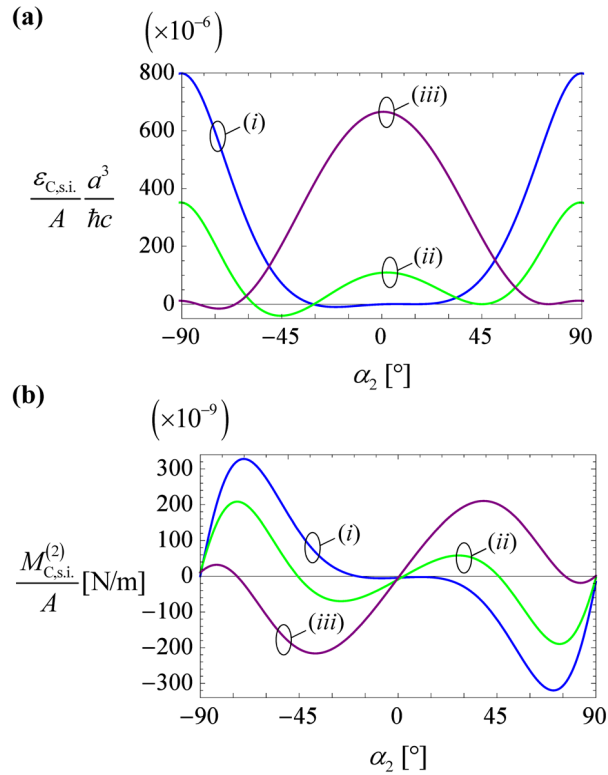

**Fig. 4.** (Color online) Normalized single-interface energy (a) and torque acting on the material 2 (b) at the junction between two identical anisotropic semi-infinite material regions (with anisotropy ratio $\chi = 10$), as a function of the angle $\alpha_2$ for a fixed $\alpha_1$. The anisotropic materials are characterized by the following parameters: $\varepsilon_t = 1$, $\omega_e / \omega_0 = 3$. (*i*) $\alpha_1 = 15°$; (*ii*) $\alpha_1 = 45°$; (*iii*) $\alpha_1 = 75°$.

Figure 4 shows the single-interface energy and torque acting on the material 2 as a function of $\alpha_2$ for three different values of $\alpha_1$: (*i*) $\alpha_1 = 15°$, (*ii*) $\alpha_1 = 45°$, and (*iii*) $\alpha_1 = 75°$. As seen, because the reduced symmetry of the system, the single-interface energy is not an even function of $\alpha_2$ and, consequently, the single-interface torque is not an odd function of $\alpha_2$, different from the results of the previous example (Fig. 3). Figure 4(a) confirms that when $\alpha_2 = \alpha_1$, i.e. when the optical axes of the two materials are aligned, the single-interface energy $\varepsilon_{C,s.i.}$ vanishes, consistent with the fact that in such a situation the system becomes equivalent to a bulk medium. In particular, the configuration with $\alpha_2 = \alpha_1$ corresponds to a local energy minimum. Somewhat surprisingly, Fig. 4(a) shows that the system has another energy minimum which occurs approximately (but not exactly) at $\alpha_2 \approx -\alpha_1$. Indeed, for $\alpha_2 \approx -\alpha_1$ the system zero-point energy has a global minimum (considering $\alpha_1$ fixed). The two energy minima correspond to positions wherein the Casimir single-interface torque vanishes (Fig. 4(b)), and hence the system has two equilibrium positions. The single-interface torque induced in the region 2 acts to rotate the "inclusions" towards the closest equilibrium point.

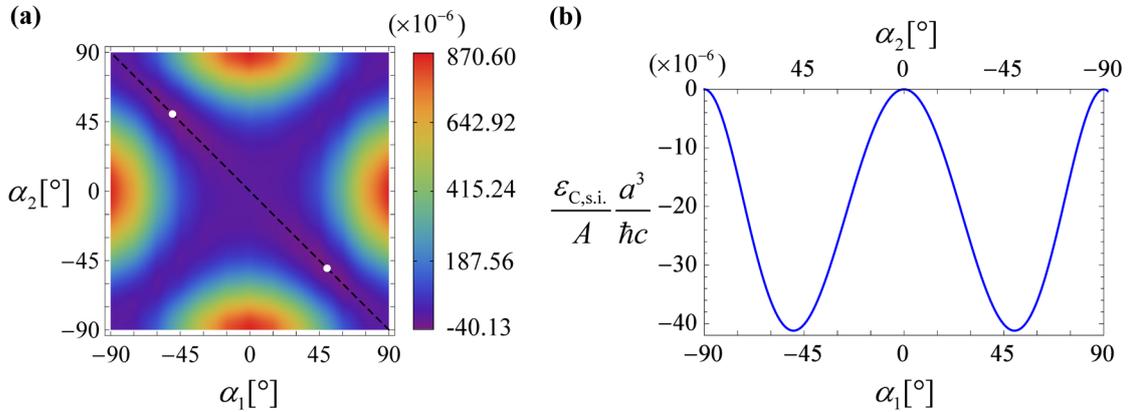

**Fig. 5.** (Color online) (a) Normalized single-interface energy $\varepsilon_{C,s.i.} a^3 / (A\hbar c)$ as a function of the orientation angles $\alpha_1$ and $\alpha_2$. The white dots represent the two global energy minima. (b) Normalized single-interface energy $\varepsilon_{C,s.i.} a^3 / (A\hbar c)$ along the line $\alpha_1 = -\alpha_2$ (black dashed line in (a)).

In order to further characterize the considered system, next it is supposed that the two particle sets are free to rotate around the *x*-axis. Figure 5(a) shows a density plot of the single-interface energy as a function of the two orientation angles $\alpha_1$ and $\alpha_2$. It can be checked that because the two materials are identical (apart from the orientation of the optical axes) the single-interface energy has the symmetries: $\varepsilon_{C,s.i.}(\alpha_1,\alpha_2) = \varepsilon_{C,s.i.}(\alpha_2,\alpha_1)$ and $\varepsilon_{C,s.i.}(\alpha,\alpha) = 0$.

The plot confirms that the system has a local energy minimum whenever $\alpha_2 = \alpha_1$. However, consistent with Fig. 4, the global energy minimum does not occur along the line $\alpha_2 = \alpha_1$, but rather along the line $\alpha_2 = -\alpha_1$. The detailed variation of the single-interface Casimir energy as a function of $\alpha_2 = -\alpha_1$ is shown in Fig. 5(b). Interestingly, the Casimir energy is negative along this line, and hence has a lower value than along the line $\alpha_2 = \alpha_1$ where it vanishes. The global energy minimum is reached at $\alpha_1 = -\alpha_2 = \pm 50°$ [see Fig. 5(b)]. Thus, if both sets of particles are free to rotate then the system tends to evolve to a configuration where the optical axes of the two sets of particles become approximately perpendicular to each other. It is important to underline that this conclusion assumes that (*i*) the torque in the bulk region vanishes ($M_{bulk} = 0$) so that all the orientations of the inclusions are equivalent in the bulk region, and (*ii*) all the dipoles in the same material region are constrained to be aligned. The physical interpretation of these results is discussed in the next subsection.

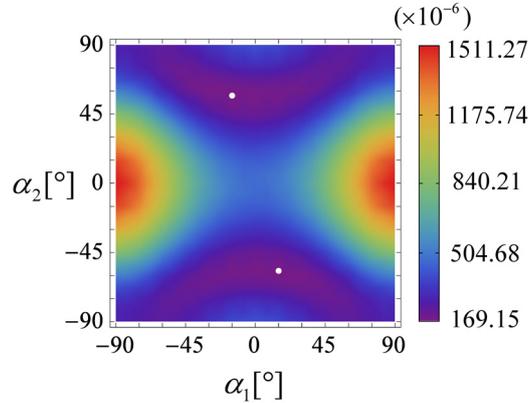

**Fig. 6.** (Color online) Normalized single-interface energy for an interface between two different anisotropic semi-infinite material regions (region 1 with anisotropy ratio $\chi = 5$ and region 2 with anisotropy ratio $\chi = 10$), as a function of the angles $\alpha_1$ and $\alpha_2$. The anisotropic material 1 is described by the parameters $\varepsilon_t = 1$, and $\omega_e/\omega_0 = 2$ whereas the anisotropic material 2 is described by the parameters $\varepsilon_t = 1$, and $\omega_e/\omega_0 = 3$. The white dots represent the two global energy minima.

A similar trend is observed when the two anisotropic materials are different. Figure 6 shows the single-interface Casimir energy for a system formed by an anisotropic material with anisotropy ratio $\chi = 5$ (region 1) and a material with anisotropy ratio $\chi = 10$ (region 2).

Now the global energy minima occur for $(\alpha_1, \alpha_2) \approx (15º, -57º)$ and $(\alpha_1, \alpha_2) \approx (-15º, 57º)$, and similar to the previous example the configurations with $\alpha_1 = \alpha_2$ are not the most energetically favorable.

To conclude this subsection, we note that for any system with the generic geometry considered in this section the single-interface Casimir energy has the following symmetries:

$$\varepsilon_{C,s.i.}(\alpha_1, \alpha_2) = \varepsilon_{C,s.i.}(\alpha_1 + \pi, \alpha_2) = \varepsilon_{C,s.i.}(\alpha_1, \alpha_2 + \pi), \tag{16a}$$

$$\varepsilon_{C,s.i.}(\alpha_1, \alpha_2) = \varepsilon_{C,s.i.}(-\alpha_1, -\alpha_2). \tag{16b}$$

The first property is trivial and is a simple consequence that the system is unchanged if the optical axis of the uniaxial dielectrics is rotated by 180º. The second property is a consequence of the fact that the zero-point energy is unaffected by a transformation of the type $y \to -y$. These properties imply that the torques must vanish ($M_{C,s.i.}^{(1)} = M_{C,s.i.}^{(2)} = 0$) when either $(\alpha_1, \alpha_2) = (0º, 0º)$ (both optical axes perpendicular to the interface) or $(\alpha_1, \alpha_2) = (90º, 90º)$ (both optical axes parallel to the interface), consistent with the numerical simulations of the previous examples [see Fig. 4(b)].

## B. Double-interface configurations

It is interesting to extend the analysis of the previous section to the case of double-interface configurations (Fig. 1(b)). In the first example, we consider an anisotropic dielectric-air-isotropic dielectric system. Figure 7 shows the magnitude of the single-interface torque and of the total torque acting on the anisotropic material as a function of $\alpha_1$ for different values of the air gap thickness ($d$). The total Casimir torque ($M_{C,tot.} = M_{C,12} + M_{C,int}$) varies considerably with the distance, since it depends not only on the single-interface torque $M_{C,12}$ (which is distance independent), but also on the interaction torque. The interaction torque $M_{C,int} = -\partial \delta \varepsilon_{C,int} / \partial \alpha_1$ is computed using Eq. (5).

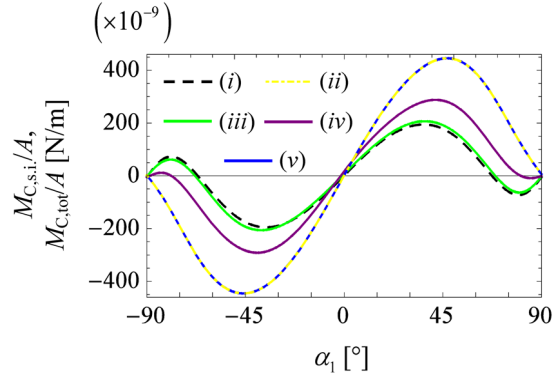

**Fig. 7.** (Color online) Torque acting on the anisotropic material as a function of the angle $\alpha_1$ for different systems. (*i*) single-interface torque for an anisotropic dielectric – isotropic dielectric interface ($M_{C,13}$), (*ii*) single-interface torque for the anisotropic dielectric–air interface ($M_{C,12}$), and (*iii*)-(*v*) total torque ($M_{C,\text{tot.}} = M_{C,12} + M_{C,\text{int}}$) for an anisotropic dielectric – air – isotropic dielectric system for different values of the air thickness $d$. (*iii*) $d = 0.01a$; (*iv*) $d = 0.1a$; (*v*) $d = 10a$. The anisotropic material has the parameters: $\varepsilon_t = 1$, $\omega_e/\omega_0 = 3$, $\chi = 10$. The isotropic dielectric ($\varepsilon_t = \varepsilon_{zz}$) has static permittivity $\varepsilon_{\omega=0} = 2$ and has $\omega_e = \omega_0$.

Importantly, Fig. 7 confirms that when $d$ tends to zero, the total Casimir torque approaches the value of the single-interface torque associated with an anisotropic dielectric - isotropic dielectric interface [see curves (*i*) and (*iii*)], as it should. Note that different from the twin-interface configurations, in systems wherein $\bar{\varepsilon}_1 \neq \bar{\varepsilon}_3$ the total Casimir energy of the system does not vanish when the gap $d$ is closed, but instead it converges to the value of the single-interface energy $\varepsilon_{C,13}$. This property confirms that the torque computed with Eq. (14) is coincident with the torque given by $M_{C,\text{s.i.},13}^{(1)} = -\frac{\partial}{\partial \alpha_1} \varepsilon_{C,\text{s.i.},131}$, ensuring that the theory is self-consistent. On the other hand, as $d \to \infty$ the total torque approaches $M_{C,12}$, i.e. the torque for a single anisotropic dielectric – air interface [see curves (*ii*) and (*v*)].

In the second example, we consider an anisotropic dielectric-anisotropic dielectric-air configuration. We assume that the two anisotropic materials, apart from the orientation of the optical axes, are identical and have the anisotropy ratio $\chi = 10$.

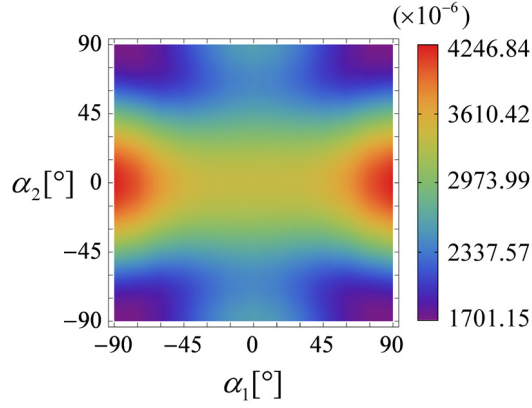

**Fig. 8.** (Color online) Normalized total Casimir energy $\varepsilon_{C,\text{tot.}} a^3/(A\hbar c)$ ($\varepsilon_{C,\text{tot.}} = \varepsilon_{C,12} + \varepsilon_{C,23} + \delta\varepsilon_{C,\text{int}}$) in a anisotropic dielectric–anisotropic dielectric–air system, as a function of the orientation angles $\alpha_1$ and $\alpha_2$. The anisotropic materials are identical and have $\varepsilon_t = 1$, $\chi = 10$ ($\omega_e/\omega_0 = 3$). The thickness of the anisotropic middle region is $d = 4a$.

Figure 8 depicts the total Casimir energy ($\varepsilon_{C,\text{tot.}} = \varepsilon_{C,12} + \varepsilon_{C,23} + \delta\varepsilon_{C,\text{int}}$) as a function of the two orientation angles $\alpha_1$ and $\alpha_2$. The density plot shows that the global energy minimum is reached when $\alpha_1 = \pm\alpha_2 = \pm 90°$, i.e., when the particles in both anisotropic material layers are parallel to the interfaces. Therefore, the air region serves to anchor the anisotropic particles of *both* anisotropic material regions.

As previously seen (Fig. 5), for a single-interface configuration formed solely by the two anisotropic materials, i.e., in the absence of the air layer, the global energy minimum is not attained when the two dipole sets are aligned. Indeed, the simulations of Fig. 5 suggest that if all the *individual* particles are free to rotate (a case which can be studied using our analytical framework) then the most energetically favorable configuration is not reached when the particles are all aligned along the same direction, but likely when they are "randomly" oriented. In other words, in a bulk material there is no "anchor" to fix a preferred alignment direction and hence unconstrained particles tend to be oriented in a "random" fashion. In contrast, the presence of the air region promotes the direction parallel to the interface as the most favorable from an energetic point of view.

## V.   CONCLUSION

We studied the zero-temperature Casimir single-interface torque at the junction between different isotropic and anisotropic materials using both microscopic and macroscopic formulations. The single-interface torque arises due to the quantum fluctuations associated with interface-type (both localized and extended) modes. These quantum fluctuations originate internal material stresses that act to change the internal configuration of the materials, i.e. to rotate the particles. The single-interface torque is quite different from the more familiar Casimir interaction torque, which is determined by the interaction of two rigid bodies separated by an isotropic material. Relying on a microscopic theory, it was proven that, in general, the single-interface torque may have a "bulk" (volumetric) contribution and a surface contribution. The torque surface component can be written in terms of the interaction energy of the system for a twin-material configuration. It was shown that the single-interface torque can be as well computed using the effective medium approximation. However, since the single-interface torque is determined by interactions of bodies that are almost in contact the use of effective medium methods is only approximately satisfactory and requires the use of a physical wave vector cut-off.

Our numerical results obtained with the continuum approximation demonstrate that in isotropic-anisotropic material systems the isotropic region acts as an anchor, imposing a preferential orientation for the particles of the anisotropic material. In particular, when the isotropic region is the vacuum the global energy minimum is reached when the dipoles are parallel to the interface. For conventional dielectrics the energy minimum moves towards the normal direction. On the other hand, in anisotropic-anisotropic material systems the global energy minimum does not correspond to a configuration with aligned dipoles, and in some cases – most remarkably when the two materials are identical – it is reached when they are approximately perpendicular. This property suggests that if all the dipoles were unconstrained and free to rotate then the configuration associated with the global energy minimum would correspond to some amorphous (non-periodic) structure with the dipoles oriented in a "random" fashion.

In future work, it will be relevant to calculate the single-interface torque with the rigorous microscopic model to assess the accuracy of the effective medium approximation. To conclude, we point out that perhaps some of the ideas discussed in this article can be experimentally validated by characterizing interfaces of liquid crystals and conventional dielectrics.

## Appendix A: Properties of the reflection matrices

In this Appendix, we derive some useful properties of the reflection matrices $\underline{\mathbf{R}}_L$ and $\underline{\mathbf{R}}_R$ for two anisotropic dielectric semi-spaces modeled as a continuum. Without loss of generality, it is supposed that the interface is normal to the $z$-direction.

To begin with, we define the transverse fields as:

$$\mathbf{E}_T = \begin{pmatrix} E_x \\ E_y \end{pmatrix} \quad \text{and} \quad \mathbf{J} \cdot \mathbf{H}_T = \begin{pmatrix} 0 & 1 \\ -1 & 0 \end{pmatrix} \begin{pmatrix} H_x \\ H_y \end{pmatrix} = \begin{pmatrix} H_y \\ -H_x \end{pmatrix}. \tag{A1}$$

Let us introduce admittance matrices such that for plane waves propagating along the $+z$ and $-z$ directions one has:

$$\mathbf{J} \cdot \mathbf{H}_T^+ = \mathbf{Y}^+ \cdot \mathbf{E}_T^+, \qquad \mathbf{J} \cdot \mathbf{H}_T^- = -\mathbf{Y}^- \cdot \mathbf{E}_T^-. \tag{A2}$$

In general, $\mathbf{Y}^{\pm}$ depend on the considered material (and in particular on the orientation of the optical axis), on the frequency $\omega$, and on the transverse wave vector $\mathbf{k}_\parallel$.

Let us consider a twin-interface configuration of the type 1-2-1. Imposing the continuity of $\mathbf{E}_T$ and $\mathbf{J} \cdot \mathbf{H}_T$ at the interfaces it is easily found that for right and left incidence the electric field reflection coefficients are:

$$\mathbf{R}_{R,121} = \left(\mathbf{Y}_2^- + \mathbf{Y}_1^+\right)^{-1} \cdot \left(\mathbf{Y}_2^- - \mathbf{Y}_1^+\right), \qquad \mathbf{R}_{L,121} = \left(\mathbf{Y}_2^+ + \mathbf{Y}_1^-\right)^{-1} \cdot \left(\mathbf{Y}_2^- - \mathbf{Y}_1^-\right). \tag{A3}$$

When $\mathbf{Y}_2^- \neq \mathbf{Y}_2^+$ the order of the matrices in the product cannot be changed.

For media invariant under inversion (transformation $\mathbf{r} \rightarrow -\mathbf{r}$ also known as the parity symmetry), e.g. standard anisotropic dielectrics, the admittance matrices are necessarily linked as:

$$\mathbf{Y}^+(k_x, k_y) = \mathbf{Y}^-(-k_x, -k_y), \qquad \text{(parity symmetry)}. \tag{A4}$$

This implies that the two reflection matrices satisfy:

$$\mathbf{R}_{R,121}(\omega, k_x, k_y) = \mathbf{R}_{L,121}(\omega, -k_x, -k_y). \tag{A5}$$

Next, we use the fact that anisotropic dielectrics are reciprocal materials [33]. In the absence of current sources, the reciprocity theorem establishes that two arbitrary solutions of Maxwell's equations, $(\mathbf{E}', \mathbf{H}')$ and $(\mathbf{E}'', \mathbf{H}'')$, in some domain with boundary $\partial D$ satisfy [33]:

$$\oint_{\partial D} (\mathbf{E}' \times \mathbf{H}'' - \mathbf{E}'' \times \mathbf{H}') \cdot \hat{\mathbf{n}}\, ds = 0. \tag{A6}$$

Here, $\hat{\mathbf{n}}$ is a unit vector normal to the surface. Let us suppose that $(\mathbf{E}', \mathbf{H}')$ and $(\mathbf{E}'', \mathbf{H}'')$ correspond to plane waves propagating along the $+z$ direction in a bulk anisotropic dielectric. The transverse wave vectors of the two field distributions are supposed to satisfy $\mathbf{k}'_\| = -\mathbf{k}''_\| = \mathbf{k}_\|$ with $\mathbf{k}_\| = (k_x, k_y, 0)$ so that the integral over the side walls (normal to the $z$-direction) in Eq. (A6) vanishes. Then, if the optical axis of the bulk anisotropic material is not aligned with the $z$-direction, the propagation constant along $z$ of the fields $(\mathbf{E}', \mathbf{H}')$ and $(\mathbf{E}'', \mathbf{H}'')$ is different. Thus, Eq. (A6) can be satisfied only if:

$$(\mathbf{E}' \times \mathbf{H}'' - \mathbf{E}'' \times \mathbf{H}') \cdot \hat{\mathbf{z}} = 0. \tag{A7}$$

This is the same as $\mathbf{E}'_T \cdot \mathbf{J} \cdot \mathbf{H}''_T = \mathbf{E}''_T \cdot \mathbf{J} \cdot \mathbf{H}'_T$ and hence it follows that $\mathbf{E}'_T \cdot [\mathbf{Y}^+(-\mathbf{k}_\|)] \cdot \mathbf{E}''_T = \mathbf{E}''_T \cdot [\mathbf{Y}^+(\mathbf{k}_\|)] \cdot \mathbf{E}'_T$ for arbitrary transverse fields $\mathbf{E}'_T$, $\mathbf{E}''_T$. Thus, we have shown that the reciprocity property implies that:

$$\mathbf{Y}^{+}\left(\omega,-\mathbf{k}_{\|}\right)=\left[\mathbf{Y}^{+}\left(\omega,\mathbf{k}_{\|}\right)\right]^{T}, \qquad \text{(by reciprocity).} \tag{A8}$$

where the superscript "*T*" stands for matrix transposition. Using now Eq. (A4), we conclude that the reciprocity and the parity symmetry impose that:

$$\mathbf{Y}^{-}\left(\omega,\mathbf{k}_{\|}\right)=\left[\mathbf{Y}^{+}\left(\omega,\mathbf{k}_{\|}\right)\right]^{T} \qquad \text{(by reciprocity and parity).} \tag{A9}$$

## Appendix B: Properties of the characteristic function *D*

Here, we derive some useful properties of the characteristic function $D$ [Eq. (2)] in the limit $d = 0^+$. The relevant materials are treated as an electromagnetic continuum. In the limit $d = 0^+$ the propagation matrices $\underline{\mathbf{M}}_{F,B}$ are identical to the unit matrix and hence it is possible to write:

$$D_{121}\left(\omega,k_x,k_y\right) = \det\left(\mathbf{1} - \mathbf{R}_{L,121}\left(\omega,k_x,k_y\right) \cdot \mathbf{R}_{R,121}\left(\omega,k_x,k_y\right)\right). \tag{B1}$$

The subscripts identify the relevant material configuration "1-2-1". The thickness of the medium 2 is infinitesimally small ($d = 0^+$). The reflection matrices are defined as in Appendix A.

The first property is a consequence of the parity symmetry discussed in Appendix A. Using the identity $\det(\mathbf{1} - \mathbf{A} \cdot \mathbf{B}) = \det(\mathbf{1} - \mathbf{B} \cdot \mathbf{A})$ (which holds for generic matrices $\mathbf{A}, \mathbf{B}$) and Eq. (A4) it follows that:

$$D_{121}\left(\omega,k_x,k_y\right) = D_{121}\left(\omega,-k_x,-k_y\right). \tag{B2}$$

The second property follows from reciprocity of the materials and establishes that:

$$D_{121}\left(\omega,k_x,k_y\right) = D_{212}\left(\omega,k_x,k_y\right). \tag{B3}$$

In the above $D_{212} = \det\left(\mathbf{1} - \mathbf{R}_{L,212} \cdot \mathbf{R}_{R,212}\right)$ represents the characteristic equation for a 2-1-2 configuration wherein the medium 1 has infinitesimal thickness. To demonstrate the second property, we use the fact that for a generic matrix $\det(\mathbf{A}) = \det(\mathbf{A}^T)$ to write:

$$D_{121} = \det\left(\mathbf{1} - \mathbf{R}_{R,121}^T \cdot \mathbf{R}_{L,121}^T\right)$$
$$= \det\left(\mathbf{1} - \left(\mathbf{Y}_2^- - \mathbf{Y}_1^-\right) \cdot \left(\mathbf{Y}_2^+ + \mathbf{Y}_1^-\right)^{-1} \cdot \left(\mathbf{Y}_2^+ - \mathbf{Y}_1^+\right) \cdot \left(\mathbf{Y}_2^- + \mathbf{Y}_1^+\right)^{-1}\right) \quad (B4)$$

The last identity is a consequence of Eqs. (A3) and (A9). Using again the property $\det(\mathbf{1} - \mathbf{A} \cdot \mathbf{B}) = \det(\mathbf{1} - \mathbf{B} \cdot \mathbf{A})$ it follows that:

$$D_{121}(\omega, k_x, k_y) = \det\left(\mathbf{1} - \left(\mathbf{Y}_2^- + \mathbf{Y}_1^+\right)^{-1} \cdot \left(\mathbf{Y}_2^- - \mathbf{Y}_1^-\right) \cdot \left(\mathbf{Y}_2^+ + \mathbf{Y}_1^-\right)^{-1} \cdot \left(\mathbf{Y}_2^+ - \mathbf{Y}_1^+\right)\right)$$
$$= \det\left(\mathbf{1} - \mathbf{R}_{L,212} \cdot \mathbf{R}_{R,212}\right) \quad (B5)$$

This result proves the desired identity [Eq.(B3)].

**Acknowledgement:** This work was funded by Fundação para a Ciência e a Tecnologia under project PTDC/EEI-TEL/4543/2014. T. M. acknowledges financial support by Fundação para a Ciência e a Tecnologia under the fellowship SFRH / BPD / 84467 / 2012. The authors gratefully acknowledge fruitful discussions with S. I. Maslovski.